\documentclass[a4paper,11pt]{article}
\pdfoutput=1 

\usepackage{jcappub} 

\usepackage{float}

\def\beq#1{\begin{equation}\label{#1}}
\def\eeq{\end{equation}}
\def\beqa#1{\begin{eqnarray}\label{#1}}
\def\eeqa{\end{eqnarray}}

\def\Eq#1{Eq.~(\ref{#1})}

\def\myfrac#1#2{\left(\frac{#1}{#2}\right)}

\def\mycomment#1{\relax}
\title{\boldmath Globular Cluster Seeding by Primordial Black Hole Population}

\author[a,b]{A. Dolgov,}
\author[c,d]{and K.Postnov}

\affiliation[a]{ITEP,  Bol. Cheremushkinsaya ul., 25, 117218 Moscow, Russia}
\affiliation[b]{Novosibirsk State University,  Novosibirsk, 630090, Russia}
\affiliation[c]{Sternberg Astronomical Institute, Moscow M.V. Lomonosov State University,\\
 Universitetskij pr., 13,  Moscow 119234, Russia}
\affiliation[d]{Faculty of Physics, Moscow M. V. Lomonosov State University, 119991 Moscow, Russia}

\emailAdd{dolgov@fe.infn.it}
\emailAdd{kpostnov@gmail.com}

\abstract{Primordial black holes (PBHs) that form in the early Universe 
in the modified Affleck-Dine (AD) mechanism of baryogenesis should have 
intrinsic log-normal mass distribution of PBHs. We show that the parameters 
of this distribution adjusted to provide the required spatial density of 
massive seeds ($\ge 10^4 M_\odot$) for early galaxy formation and not violating the 
dark matter density constraints, predict the existence of the population of intermediate-mass PBHs with 
a number density of  $\sim 1000$~Mpc$^{-3}$. We argue that the population of intermediate-mass AD PBHs 
can also seed the formation of globular clusters in galaxies. In this scenario, each globular cluster should host 
an intermediate-mass black hole with a mass of a few thousand solar masses, and should not obligatorily be immersed in 
a massive dark matter halo.   
}

\begin{document}
\maketitle
\flushbottom

\section{Introduction}
\label{sec:intro}

Primordial black holes (PBHs) formed at the radiation-dominated epoch
in the early Universe~\cite{1967SvA....10..602Z,1974MNRAS.168..399C}
have been topical for many years as possible dark matter (DM) candidates (see, e.g., \cite{2016PhRvD..94h3504C} for the recent review).
PBH can have a very broad mass range, and there are various constraints on their
possible abundance from different astrophysical observations \cite{2016PhRvD..94h3504C,2016PhRvD..94f3530G,2017arXiv170203901G}.
Recently, the interest to PBHs has been renewed in connection with the discovery 
of gravitational waves by LIGO~\cite{LIGO-PRL} from coalescing massive ($\sim 36+29 M\odot$)  black holes.
It has been recognized that coalescing binary stellar-mass PBHs can be quite abundant sources of GWs \cite{1997ApJ...487L.139N}, 
and immediately after the LIGO GW150914 announcement, several papers proposed PBHs as viable explanation of 
the derived properties of black holes in GW150914 \cite{2016PhRvL.116t1301B,2016PhRvL.117f1101S, 2016PASJ...68...66H,2016JCAP...11..036B}. 
Importantly, as stressed in \cite{2016JCAP...11..036B}, the PBH hypothesis for LIGO GW150914 is supported by
the derived upper limits of low angular momenta of coalescing components in this source 
\cite{2016PhRvL.116x1102A}, which may be difficult to respect in the standard binary evolution scenario 
\cite{2016MNRAS.462..844K}.

In \cite{2016JCAP...11..036B}, we considered 
a specific model of PBH formation which both naturally reproduces the extreme properties of
GW150914, the rate of binary BH merging events as inferred from the first LIGO science run 9-240
Gpc$^{-3}$~yr$^{-1}$ \cite{LIGO_doc2016},
and provides seeds for early supermassive BH formation in galaxies.
The model is based on the modified Affleck-Dine~\cite{ad-bg} scenario for baryogenesis \cite{ad-js}.
The model was discussed in more details in ref.~\cite{ad-mk-nk}, applied to an
explanation of the early supermassive BH (SMBH) observations at high redshifts~\cite{ad-sb}, and to a
prediction and study of the properties of possible antimatter stellar-like objects in the Galaxy~\cite{cb-ad,sb-ad-kp}.

Different model PBH as seeds for early galaxy formation
have been discussed earlier (e.g. \cite{2001JETP...92..921R,2008ARep...52..779D,2010RAA....10..495K,2012arXiv1208.3999D}).
Here we show that the universal log-normal PBH mass distribution predicted in~\cite{ad-js}
with parameters fixed from the observed spatial density of supermassive BHs in galaxies also automatically 
satisfy all DM constraints and predicts the number density of intermediate-mass BHs consistent with 
the number density of globular clusters (GC) in galaxies. 

\section{The modified Affleck-Dine scenario and universal log-normal PBH mass distribution
\label{s-model}}

In the modified by coupling to inflaton~\cite{ad-js} 
 supersymmetric scenario for baryogenesis (the Aflleck-Dine mechanism~\cite{ad-bg})
small bubbles with huge baryon asymmetry, even of order unity, might arise at inflationary stage.
These bubbles could have astrophysically interesting sizes  but
occupy a relatively small fraction of the whole volume of the universe. Due to large baryonic number,
they could make a noticeable
contribution to the total cosmological energy density.
The shape of the mass distribution of these high-B bubbles is determined by inflation and
thus it is virtually model independent and has the log-normal form:
\begin{equation}
\frac{dn}{dM} = \mu^2 \exp\left[-\gamma\,\ln^2(M/M_{0}) \right],
\label{dn-dM}
\end{equation}
where $\mu$, $\gamma$, and $M_{0}$ are some unknown, model dependent constant parameters.

Originally, the induced small-scale baryonic number fluctuations lead predominantly to isocurvature fluctuations, but
after the QCD phase transition at a temperature of $T \sim 100 - 200$~MeV, when the massless quarks have combined into heavy
nucleons, the isocurvature fluctuations transformed into large density perturbations at
astrophysically large but cosmologically small scales.
Depending upon the history of their formation, the high-B bubbles could be turned into PBH, compact stellar-like
objects, or even rather dense primordial gas clouds. 
Note that the B-bubbles are formed predominantly spherically symmetric because such configurations minimizes
the bubble energy. Their angular momentum should be zero because they are formed as a result of phase transition in
cosmological matter with vanishing angular momentum
due to absence of the cosmological vorticity perturbations.

At the high mass tail of the AD PBH distribution, superheavy PBHs (with masses of $\sim 10^5 M_\odot$ and even higher) 
might be created \cite{2016JCAP...11..036B}.
The emergence of supermassive BHs in galaxies at redshifts of order 10 is in serious tension with the conventional mechanisms of their formation.  
In the last few years, there is a growing evidence for the presence of early supermassive BHs  
which later can serve as seeds for galaxy formation (see, e.g., sec. 5 of ref.~\cite{ad-hi-z} or more recent review in the lecture~\cite{ad-itep-16}).

The initially formed superheavy AD PBH might have much smaller masses
(around $10^4-10^5 M_\odot$) to 
subsequently grow to $10^{9} M_\odot$ because of an efficient accretion of matter and
mergings (see the state-of-the-art SMBH growth calculations in \cite{2016MNRAS.462..190R}). This mass enhancement factor is
much stronger for heavier BH and
thus their mass distribution may be different from (\ref{dn-dM}). However, we assume that the original BHs were created with
the universal distribution (\ref{dn-dM}).

\section{Intermediate-mass AD PBHs}

Let us consider the model parameters $M_0$ and $\gamma$ of the universal AD PBH mass distribution 
(\ref{dn-dM}). In natural units $\hbar=c=1$, the solar mass is $1 M_\odot\approx 1.75\times 10^{95}$~Mpc$^{-1}$, 
the normalization constant $\mu = 10^{-43}$ Mpc$^{-1}$. With this values, for example, setting $\gamma = 0.5$, $M_{\max} =
M_\odot$ yields the total energy density of AD PBHs a few times smaller than the total mass density of matter 
(both DM and baryonic), $\rho_m\approx 4\times 10^{10}M_\odot$~Mpc$^{-3}$. Below we shall normalize mass 
to the solar mass and use dimensionless variables $x=M/M_\odot$, and $y=M_0/M_\odot$. As shown in \cite{2016JCAP...11..036B}, 
it is convenient to relate the parameter $\gamma$ and $y$ such that the fraction of AD PBHs in the mass
range $0.1-1 M_\odot$ be $f=\rho_{BH}/\rho_m=0.1$ in the interval $\gamma=0.4-1.6$, yielding the relation 
\beq{y}
y\approx \gamma+0.1\gamma^2-0.2\gamma^3\,.
\eeq

\begin{figure}
\begin{center}
\includegraphics[scale=0.35]{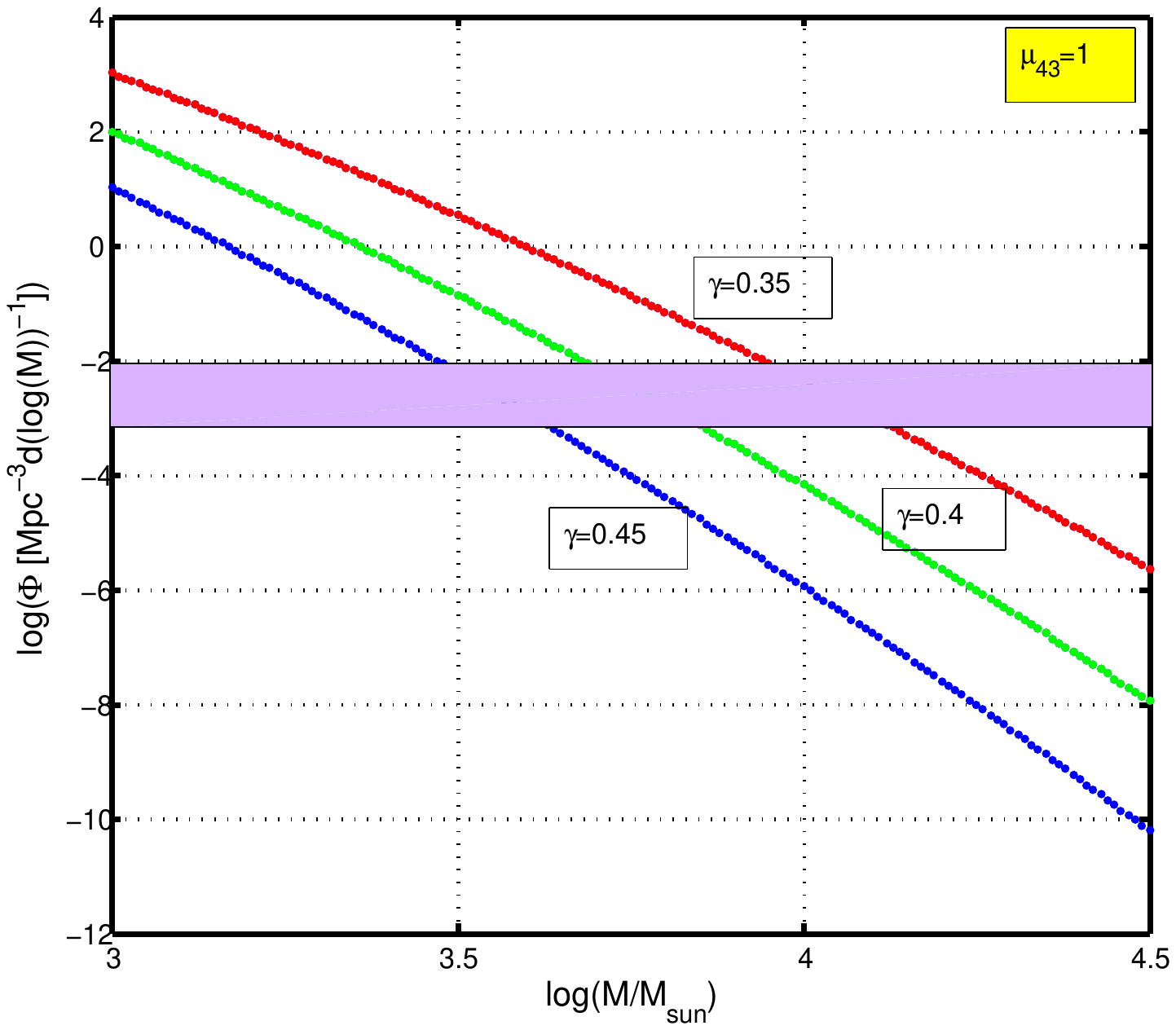}
\hfill
\includegraphics[scale=0.35]{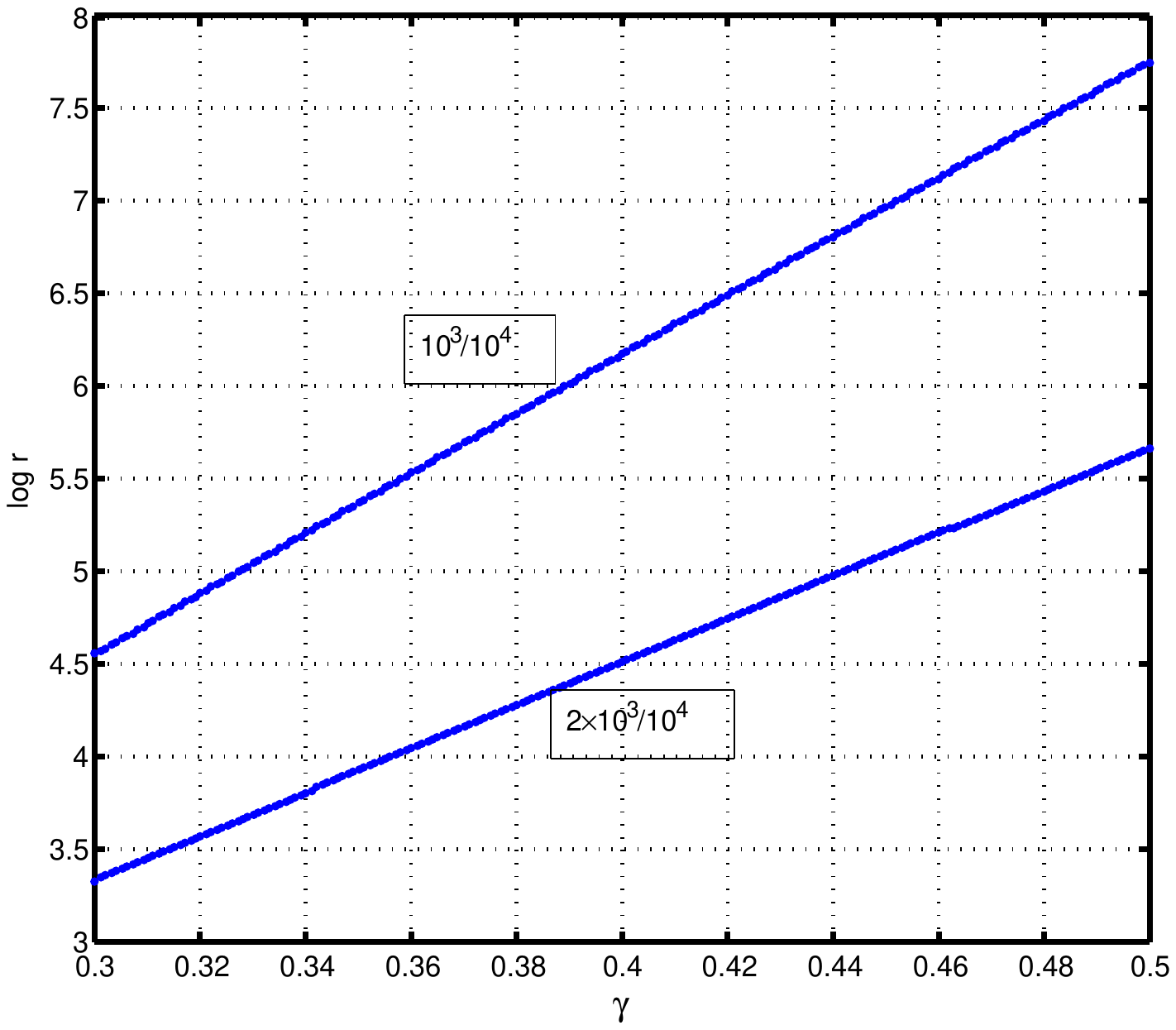}
\end{center}
\caption{Left: space density of AD PBH (per Mpc${^{-3}}$ per ${d\log M)}$  for different values of the $\gamma$ parameter, 
as compared to those derived from observations  of SMBH in large galaxies (purple rectangle). Right: the ratio of 
AD PBH number per logarithmic mass interval with mass $10^3$ and $2\times 10^3 M_\odot$ to that of AD PBH
with $M>10^4 M_\odot$ as a function of the $\gamma$ parameter. }
\label{f1}
\end{figure}

Now, the fraction of AD PBHs (per logarithmic mass interval) in the mass range around $M_1$ to that in the mass range around $M_2$
becomes 
\beq{r}
r=\myfrac{x_1}{x_2}e^{\gamma \left[ \ln(x_2/y)^2-\ln(x_1/y)^2 \right] }=
\left(\frac{x_2}{x_1}\right)^{-1+\gamma\ln \left( x_1x_2/y^2 \right)}\,.
\eeq
Consider IMBH with $2\times 10^3 M_\odot$ suspected to exist in the globular cluster cores \cite{2017Natur.542..203K}.
Assume after \cite{2016MNRAS.462..190R} that all AD PBH with $M>10^4 M_\odot$ were seeds to SMBHs in galaxies which grew up to SMBH presently observed. 
For the case of interest, $x_2=10^4>x_1=2\times 10^3\gg 1$, and therefore 
\beq{}
r\approx \left(\frac{x_1x_2}{y^2}\right)^{\gamma\ln(x_2/x_1)}\,,
\eeq
or 
\beq{logr}
\log r\approx  \gamma \log e \,\log(x_2/x_1)\log(x_1x_2/y^2)\,.
\eeq
Noticing that for $\gamma<1$ we have from (\ref{y}) $y\simeq \gamma$, we can solve  \Eq{logr} to find 
$\gamma\approx 0.4$ for $r=10^5$. The plot of $\log r$ from Eq. (\ref{r}) as a function of $\gamma$ is shown in Fig. 
\ref{f1} (right panel). It is seen from this figure that for parameter $\gamma\sim 0.35-0.45$, the number of AD PBH with masses 
$(2-3)\times 10^3 M_\odot$ is about $10^4-10^5$ per one AD PBH with mass $>10^4 M_\odot$. If all AD PBH with initial 
mass $>10^4 M_\odot$  turned into SMBHs in galaxy nuclei during the evolution of the 
Universe with the present-day space density of SMBH about $10^{-2}-10^{-3}$ per cubic Mpc (purple rectangle in the 
left panel of Fig. \ref{f1}), the present-day AD IMBH space number density would fall into interval 
$\sim 10^2-10^3$ per cubic Mpc. 
This density of AD IMBHs is sufficient to seed the formation of globular clusters in galaxies (see the next Section).

\section{IMBH as globular cluster seeds}

The abundance of primordial IMBH in the AD scenario obtained above ($\sim 10^5-10^6$ per present-day SMBH) may have interesting implications to the 
formation of the globular clusters (GC) which for a long time are though to be the oldest stellar systems in the Universe
\cite{1968ApJ...154..891P,1984ApJ...277..470P}. Presently, GCs are known to have no extended dark halos \cite{1996ApJ...461L..13M,2016MNRAS.462.1937S}. Their formation 
mechanisms and evolution involve many uncertainties and are actively studied (see e.g. \cite{2015MNRAS.454.1658K,2015MNRAS.454.4197R,2016MNRAS.462.2861R} and references therein).

 In the concordance hierarchical $\Lambda$CDM cosmology GCs can be formed within dense gas discs during early galaxy formation at high redshifts $12>z>3$ \citep{2005ApJ...623..650K}. However, the presence of primordial IMBHs may change this picture.
A possible scheme can be as follows. A primordial IMBH creates gravitational potential $\phi_{BH}=-GM_{BH}/r$. According to the standard 
picture of structure formation in the hierarchical cosmology \citep{2001PhR...349..125B,2013RPPh...76k2901B}, baryonic mass that can accrete into the potential well $\phi$ is of order of the dark halo mass that creates this potential well and is of order of the cosmological Jeans mass
\beq{}
M_b\sim M_{vir}\sim M_J\approx 10^4M_\odot\myfrac{1+z}{10}^{3/2}\,.
\eeq
In this picture, at redshifts $z\gtrsim 20$ halos with masses $<10^4M_\odot$ never collapses since the 
virial gas temperature is below $T_{CMB}$. 

The present-day AD IMBH space density of $\sim 10^2-10^3$ per cubic Mpc means that at $z\sim 10$ their 
space number density could be as large as $n_{IMBH}\sim 10^5-10^6$~Mpc$^{-3}$, comparable to the 
expected abundance of dark matter halos  with masses $10^3-10^4 M_\odot$  at $z\sim 10$ in the simplest hierarchical structure formation 
scenario based on the Press-Schechter formalism \citep{2001PhR...349..125B}. That is, some IMBHs can sink into dark
matter potential well. We remind that 
AD IMBH form practically independently of the production of the primordial  perturbation spectrum generated 
by inflation, however, their average energy density may be slightly inhomogeneous and trace 
the usual density fluctuations induced at the end of inflation. 
 Clearly, 
the growth of non-linear structures in the presence of AD IMBH deserves further studies.

The IMBHs either can merge with CDM halos, or may remain as isolated, DM-free
potential wells filled with hot gas that collapse at smaller redshifts, at the reionization stage and galaxy formation epoch,
and can be seeds for globular clusters around galaxies. 
Current observations indeed discover proto-GCs at redshift $z>3$ \citep{2017MNRAS.467.4304V}. 

The simplified estimates given above suggest that baryonic structures with mass similar 
to that of GCs can be formed around seed IMBHs without help of dark mater halos at the stages preceding the early galaxy formation. 
Of course, the real picture should be much more complicated due to the parallel development of non-linear dark matter structures, baryon cooling, etc., which requires numerical simulations.  

In the discussed here scenario of GC formation with primordial IMBH seeds, a GC can 
initially host a central IMBH and is not obligatory immersed in a massive dark matter halo. While observational detection of IMBHs in GC cores is challenging, there are signatures that indeed IMBHs with masses up to a few $10^4 M_\odot$ can be present in GC cores \cite{2017Natur.542..203K,2017MNRAS.464.2174B}.


The central IMBH in a GC could form differently from the scenario we discuss, 
for example from primordial Population III star evolution \cite{2001ApJ...551L..27M} or as a result of runaway stellar collisions of massive stars in young star clusters \cite{2004Natur.428..724P}. Here more observations are required 
to disentangle different IMBH formation scenarios (see \cite{2016MmSAI..87..555G} for a brief review of other IMBH formation mechanisms).

The presence of IMBH in the GC cores can substantially affect the GC evolution \cite{2013A&A...558A.117L}.  
Note in this respect that the universal relation between
the central BH mass and bulge mass in galaxies \cite{2013ARA&A..51..511K} could be also applied to GC IMBHs taking into account the
decrease of GC mass during the evolution (especially due to tidal stripping) \cite{2013MNRAS.434L..41K}. 
In addition to GCs hosting IMBHs, recently discovered dark star clusters with high mass-to light ratio are though to harbor central IMBHs \cite{2015ApJ...805...65T,2016ApJ...832...88B} and could be the remnants of tidally disrupted dwarf spheroidals having masses intermediate between GCs and large galaxies. 

We conclude that primordial AD IMBHs with mass of a few thousand solar masses can be important additions to the standard paradigm of the 
early structure formation in the hierarchical cosmogony \cite{2001PhR...349..125B}. In the proposed scenario, all PBHs with masses in excess of 
$\sim 10^4 M_\odot$ assembled into SMBHs presently observed in centers of galaxies, while PBHs with smaller masses served as seeds for GC cores which may be not surrounded by their own dark matter halo. If this is correct, GCs should generically host the central IMBH, which can be tested in future observations. 

\acknowledgments

The authors thank A.V. Zasov, O.K.Sil'chenko and A.S.Rastroguev for useful discussions and the referee
for critical remarks. 
A.D. acknowledges support of the Grant of President of Russian Federation
for the leading scientific Schools of Russian Federation,
NSh-9022-2016.2. K.P. acknowledges support from RSF grant 16-12-10519 (the analysis of
PBH as seeds for globular clusters).

\bibliography{puzzlesBH}

\bibliographystyle{JHEP}

\end{document}